\documentclass[aip,jcp,reprint]{revtex4-1}

\usepackage{graphicx}
\usepackage{amsmath,amssymb}

\begin{document}

\title{Glass-like dynamics of the strain-induced coil/helix transition on a permanent polymer network}
\author{O. Ronsin}
\author{C. Caroli}
\author{T. Baumberger}
\affiliation{Institut des NanoSciences de Paris, CNRS, Sorbonne Universit\'e -- Pierre et Marie Curie, UMR  7588, 4 place Jussieu, 75005 Paris, France}

\begin{abstract}
We study the stress response to a step strain of covalently bonded gelatin gels in the temperature range where triple helix reversible crosslink formation is prohibited. We observe slow stress relaxation towards a $T$-dependent finite asymptotic level. We show that this is assignable to the strain-induced coil $\rightarrow$ helix transition, previously evidenced by S. Courty, J.L. Gornall and E.M. Terentjev (PNAS, {\bf 102}, 13453 (2005)), of a fraction of the polymer strands. Relaxation proceeds, in a first stage, according to a stretched exponential dynamics, then crosses over to a terminal simple exponential decay. The respective characteristic times $\tau_{K}$ and $\tau_{f}$ exhibit an  Arrhenius-like $T$-dependence with an  associated energy $\cal E$  incompatibly larger than the activation barrier height for the isomerisation process which sets the clock for an elementary coil $\to$ helix transformation event. We tentatively assign this glass-like slowing down of the dynamics to the long-range couplings due to the  mechanical noise generated by the local elementary events in this random elastic medium.

\end{abstract}

\maketitle

\section{Introduction}
The slow relaxation dynamics of glass-forming deeply supercooled liquids exhibits, following a thermal or mechanical quench, well-known characteristics \cite{AngellMcK}. It is in general well described by the phenomenological Kolrausch expression, namely a stretched exponential $\sim \exp[-\left( t/\tau_{\alpha}\right)^{\beta}]$, with $0 < \beta < 1$.  The $\alpha$-relaxation time $\tau_{\alpha}$ exhibits, for a vast majority of materials, the so-called  fragile behavior, {\it i.e.} a faster than Arrhenius growth as temperature $T$ decreases towards the glass transition temperature $T_g$ -- a behavior which can be formulated as the paradoxical increase upon cooling of an "apparent activation energy".

It is now recognized that the elementary relaxation events responsible for this dynamics are rearrangements of clusters of a few (typically $\leq 10$) structural units (atoms, monomers,...), which involve coordination changes  \cite{Dyre}. Each such event thus necessarily gives rise to a long range elastic field which propagates into the deformable embedding solid, as evidenced in the simulation results of \cite{PRLAnael}. How and to which extent the collective effects of the interplay between this "dynamical noise" and the elastic non-affinity due to structural disorder result in the spectacular slowing down observed upon approaching $T_g$ remains for the moment an essentially open issue. 

Here, in order to try and shed some further light on this question, building upon recent results of Terentjev and coworkers \cite{Kutter,Courty}, we perform an experimental study of the stress evolution triggered by an applied strain in a  highly deformable permanent elastic network, obtained by covalent cross-linking of biopolymer (gelatin) chains. In this system, relaxation proceeds via local transformation events coupled together by the stress noise they generate in the network on which they take place. As cross-links (CL) are permanent, the network topology remains invariant: in contrast with the case of glass-formers, relaxation does not affect coordination. Note that the associated decimation of relaxation paths in the energy landscape is reminiscent  of the effect of selective particle pinning in studies of point-to-set correlations in glass-forming liquids \cite{PointtoSet}.

The experimental procedure consists in applying to this elastic solid a step shear strain which, as demonstrated in \cite{Courty}, triggers, on the stretched polypeptidic inter-crosslink strands, the 
coil$\to$helix transition of successive monomer segments. It is these elementary {\it cis-trans} isomerisation processes which play the role of relaxation-promoting local events. Any such process induces a variation of the tension of the strand on which it occurs, hence a set of stress signals in the embedding network, which constitute the dynamical noise. We follow the response of the system by monitoring the relaxation of the macroscopic shear stress  $\sigma$. 

We observe a dynamics characterized by the following features :

$\bullet$  Relaxation of $\sigma$ towards a finite asymptotic value. The associated stress drop $\Delta \sigma$ grows non linearly with the applied strain $\gamma_0$. At fixed $\gamma_0$, it decreases with T, until the effect vanishes above a threshold temperature.  This behavior is fully consistent with, and thus confirms, the picture put forward by Terentjev et al. of the role of the strain-induced coil $\to\alpha$-helix transition in the mechanical response of networks of denatured polypeptides. 
   
$\bullet$ The relaxation dynamics itself exhibits two distinct stages: the earlier one is well fitted by a Kolrausch stretched exponential expression. From this we identify a relaxation time $\tau_K$, many orders of magnitude larger than the Rouse time of the inter-CL strands, which is here the relevant viscoelastic time, since the CL are irreversible and the gelatin concentration is such that our initial polymer solution is in the non-entangled regime. 
      
This early stretched exponential response crosses over to a simple exponential decay towards the above-mentioned asymptotic stress drop $\Delta \sigma$, thus defining a terminal relaxation time $\tau_{f}$, which turns out to be comparable with the cross-over time $\tau_{co}$.
   
We find that, in the limited temperature range (see Section II) accessible to our experiments, 
$\tau_{K}(T)$ and $\tau_f(T)$ both exhibit an Arrhenius-like behavior, with  the same activation energy ${\cal E}$. However, ${\cal E}$ turns out to be much larger than that, $E_{act}$, for the prolyl cis-trans isomerisation process which has been long ago recognized to set the clock for the coil $\to\alpha$-helix transition involved in the renaturation of gelatin into collagen \cite{Bachinger}. 

On the basis of the observation of {\it(i)} stretched exponential relaxation on time scales way too large to be assignable to simple network viscoelasticity and {\it(ii)} an apparent Arrhenius slowing down upon cooling much larger than expected from that of the elementary microscopic relaxation events,
we are led to conclude that the elastic noise generated by local relaxation events is sufficient to bring the response to quench of a solid exhibiting frozen structural disorder into the class of glass-like slow dynamics.  

\section{Experimental}
The system we study is a gelatin hydrogel. Gelatin is obtained from collagen, whose molecules are constituted of the wrapping into a right-handed helix of three left-handed single strand  $\alpha$-like (polyproline II) helices. Collagen denaturation results, at high temperatures (typically $\gtrsim 30^{\circ}$C), in solutions of single stranded chains in the coil configuration. 
  
Following the pioneering study by Flory and Weaver \cite{Flory}, considerable effort has been devoted to analyzing the details of the kinetic path leading to collagen renaturation \cite{Harrington}. It is now agreed that the kinetics is limited by {\it cis-trans} isomerisation of the prolyl peptide bonds present in large proportion along the chains. The associated activation energy $E^{0}_{act} \simeq 72$ kJ/mol is quite large, so that the clock-setting time for the dynamics of the coil to helix transition $\tau_{c/h}$ lies in the unusually slow range of a few ten seconds \cite{Engel}.  

This transition is in general triggered by cooling. Besides, several theoretical works \cite{Buhot,Pincus}     have predicted that imposing a finite end-to end extension $R$  to a molecule of a polymer exhibiting a coil/helix transition results in a displacement of the transition at the expense of the coil state. Indeed, as $R$ is increased, due   to the decrease of the coil entropy, the free energy balance is increasingly biased in favor of the helix state. 
So, for example, at a temperature $T$ slightly larger than the free chain transition one $T_{0}$, one expects a finite helix fraction to appear beyond  some extension threshold which decreases as $T$ is lowered. This fraction is predicted to grow with $R$ until the molecule becomes fully helical, beyond which further extension results in its decrease. Such effects have in particular a bearing on single molecule force-extension curves \cite{Buhot}. 

More recently, Kutter and Terentjev \cite{Kutter} have extended a simplified version of the theory  to the case of polymer networks. In this latter case, since the end-to-end distances of the polymer strands are imposed by neighboring crosslinks, submitting a gel  to a mechanical deformation must induce both a variation of the helix content and a related contribution to the stress response. Courty, Gornall and Terentjev \cite{Courty} then measured the optical activity of stretched thin plates of gelatin gels, which gives access to the helix content, and showed that its non-monotonic dependence on the strain amplitude is in good qualitative agreement with their theoretical predictions. 

For our present purpose, it  is important to note that the system used in these experiments was a {\it physical gel}, obtained by cooling  a gelatin-in-water solution below $T_{gel} \sim 30^{\circ}$C. Under such conditions, gelation occurs via the  formation of  segments of the original triple helix collagen structure interconnected by coiled strands. 
The H-bonds which stabilize these cross-links (CL) are weak enough for  the gels to be thermoreversible. That is, as the CL are able to rearrange via partial zipping/unzipping of the triple helix under the combined effect of internal and applied stresses, gelatin gels obtained by cooling exhibit slow relaxation and glass-like aging features, such as logarithmic shear modulus strengthening  \cite{Normand, Nij}, sensitive to applied stress \cite{NousAging}.

As developed in Section I, we are interested here in studying the dynamics of strain-induced relaxation in a network where monomer "transfer" between contiguous strands is prohibited. This makes physical gels unsuitable, due to  the fact that biased zipping/unzipping at the ends of a triple helix segment results in strand sliding. We circumvent this difficulty by making use of gels resulting from the covalent bonding, via an enzymatic route, of gelatin chains in their high temperature coil state. We then study, in the temperature range where physical triple-helix CL formation does not occur, the full time-dependent stress response to a step strain.

\subsection{Gel preparation and characterization}
In order to prepare covalently bonded gelatin networks with a given shear storage modulus $G$, we proceed as described  in detail in\cite{GelZ}. In short, we dissolve gelatin (300 Bloom, type A from porcine skin, Sigma) in deionized water et 65$^\circ$C. After total dissolution of the polymer,  the solution is quickly mixed  at 40$^\circ$C with a Tgase enzyme solution (microbial transglutaminase, Activa-WM, Ajinomoto Foods Europe SAS) so as to reach a final composition of 5 wt\% gelatin and 2.6 nmol of Tgase (corresponding to an enzymatic activity of 2U). 
We have checked that, for this concentration, the gelatin solution is in the semi-dilute, non entangled regime\cite{Supp}. The solution is then poured into the temperature-controlled cell of a stress-controlled rheometer (MCR 501, Anton Paar) equipped with a cone-plate, sand-blasted cell. The sample is protected against solvent evaporation  by a paraffin oil rim. The cell temperature is controlled to within 0.1$^\circ$C with the help of a thermoelectric device. 

\begin{figure}[htbp]
\begin{center}
\includegraphics[width = 8.5 cm]{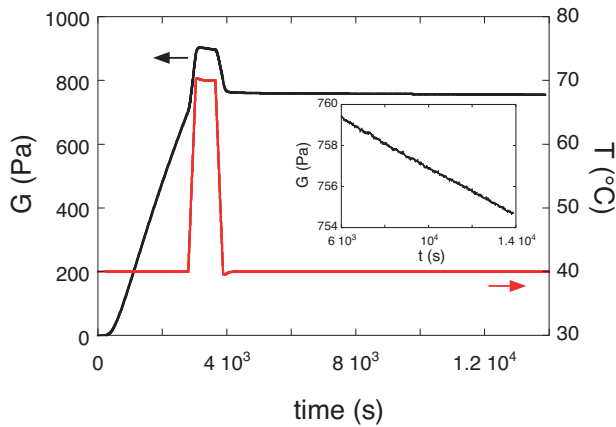}
\caption{Build-up of the shear modulus $G(t)$ (dark curve) in response to the thermal history (light red curve). Inset: Blow-up  of the late stage of shear modulus evolution.}
\label{fig:preparation}
\end{center}
\end{figure}

Gelation then proceeds at $T_{set} = 40^\circ$C. At this temperature, chosen well above $T_{gel} \simeq 30^\circ$C, no triple-helix {\it reversible} cross-link can form whereas Tgase catalyzes actively  the formation of inter-chain {\it covalent} bonds between two specific residues.  
The advancement of the cross-linking process  is monitored by measuring the storage shear modulus $G$ at $f =1$ Hz with a $1\%$ strain amplitude.  When the target $G$ level is reached, we quickly heat the sample up to 70$^\circ$C and maintain it at this temperature for 10 min, after which it is cooled down back to the working temperature $T>T_{gel}$. After such a heating stage, the enzyme is known to be inactivated, i.e. further covalent cross-linking is fully inhibited. Indeed, as illustrated on Figure \ref{fig:preparation}, the gel shear modulus no longer increases.

However, closer inspection (see inset of Figure\ref{fig:preparation}) reveals that $G(t)$ systematically exhibits a very slow and {\it linear} decrease. This feature is clearly assignable to the presence in the enzyme preparation of traces of protease \cite{GelZ}, the effect of which is to weaken the gel by catalyzing the scission of gelatin network  strands.  This is consistent with the  
fact that (see inset of Fig.\ref{fig:raw}), the higher the temperature, the steeper the modulus decrease.
 
 \subsection{Stress relaxation experiments}
 
Stress relaxation experiments are performed according to the following protocol. After quenching  at a rate of 15$^\circ$C/min from $70^\circ$C  to the working temperature $T$, we shear the sample at the rate $0.1$ s$^{-1}$ until we reach the strain level $\epsilon =  20\%$. From this instant which we choose as the time origin, we record the shear stress signal $\sigma_{raw} (t)$. 

\begin{figure}[htbp]
\begin{center}
\includegraphics[width = 8.5 cm]{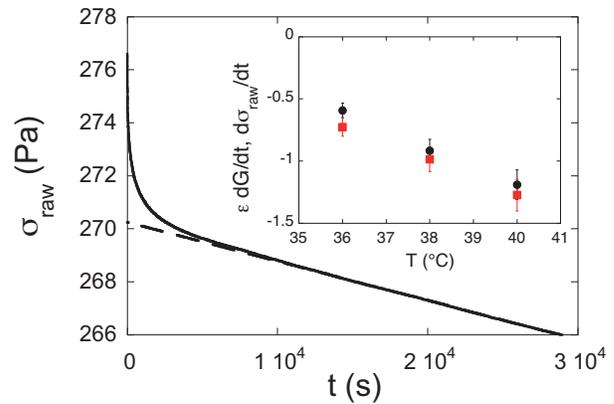}
\caption{Stress response $\sigma_{raw}(t)$ of a gel with initial shear modulus $G  = 1400$ Pa following a step strain of amplitude $\epsilon = 20\%$, at $T = 35^\circ$C. Dashed line: asymptotic linear decrease. Inset: Temperature dependences of the asymptotic slope  $(d\sigma_{raw}/dt)_{t\to\infty}$ (red squares) and of the decay $\epsilon\,dG/dt$ (black dots) due to the protease-induced modulus decrease  (vertical unit: $10^{-4}$ Pa.s$^{-1}$). The error bars correspond to the scattering of the results from three different samples.}
\label{fig:raw}
\end{center}
\end{figure}

As can be seen on Figure \ref{fig:raw}, which displays a typical recording, the stress relaxes, and exhibits a {\it linear} asymptotic decrease. On the other hand, we have seen that protease trace contamination results in a {\it linear} decrease of the gel shear modulus $G$ with time. Since the gel sample is under the constant shear deformation $\epsilon$ (which we have checked to lie within the linear elastic response regime), the network weakening obviously contributes to the decrease of the shear stress $\sigma_{raw}$ at a rate $\epsilon dG/dt$. As shown on Fig.\ref{fig:raw} (inset) this contribution fully accounts  for the asymptotic decrease of $\sigma_{raw}$, since $ (d\sigma_{raw}/dt)_{t\to\infty}$ and $\epsilon dG/dt$ are equal within experimental error. In order to correct for this linear additive drift, we define the  stress:
$$\sigma(t) = \sigma_{raw}(t) - t(d\sigma_{raw}/dt)_{t\to\infty}$$
It is this corrected, intrinsic, stress which is dealt with in the following sections.

\section{Results}

\begin{figure}[htbp]
\begin{center}
\includegraphics[width = 8.5 cm]{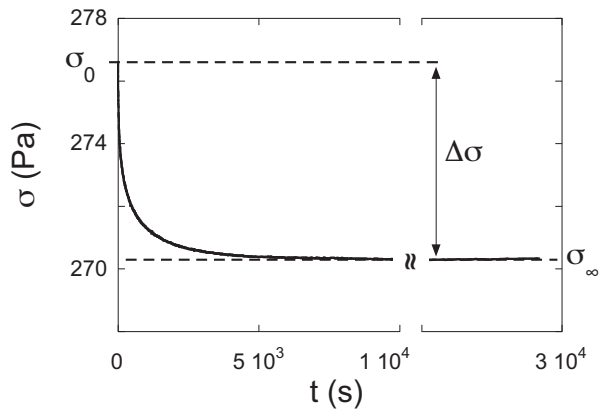}
\caption{Intrinsic stress response $\sigma(t) = \sigma_{raw}(t) - t(d\sigma_{raw}/dt)_{t\to\infty}$. Same data as Fig.\ref{fig:raw}. }
\label{fig:relax}
\end{center}
\end{figure}

Figure \ref{fig:relax} shows a typical stress response $\sigma (t)$, obtained at $T = 35^\circ$C for a gel of shear modulus $G = 1400$ Pa under the applied strain $\epsilon = 20\%$.  
After gradual relaxation extending over  thousands of seconds the stress drop  $\sigma_0-\sigma(t)$ saturates at the value $\Delta\sigma = \sigma_{0} - \sigma_{\infty}$. Note that, although   $ \Delta\sigma/\sigma_0$ lies in the $10^{-2}$ range, the stress drop value remains much larger than the noise level.

\begin{figure}[htbp]
\begin{center}
\includegraphics[width =8.5 cm]{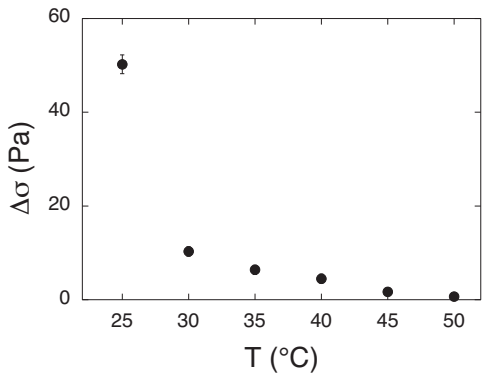}
\caption{Temperature dependence of the stress drop $\Delta \sigma = \sigma_0 -\sigma_{\infty}$ (see  Fig. \ref{fig:relax}) for gels with $G = 1400$ Pa. Except for the $T = 25^\circ$C datum, the error bars are smaller than the dots.}
\label{fig:deltasigmadeT}
\end{center}
\end{figure}

Figure \ref{fig:deltasigmadeT} displays the evolution of $\Delta \sigma$ with temperature in the range where physical cross-linking upon cooling does not occur. Note that for   $G = 1400$ Pa this range extends down to the value $T = 25^\circ$C lower than $T_{gel}$. Indeed we have checked that, as suggested by previous results \cite{Mixte}, the shear modulus of the sample remained constant over the duration ($\sim 10^5$ s) needed for full stress   relaxation at this temperature. 
$\Delta \sigma(T)$ is seen to decrease steeply down to a negligible level, reached for $T\gtrsim 45^\circ$C. 

In order to assess the degree of thermal reversibility of the physical process underlying relaxation in our system, we have performed the following control experiment\cite{Supp}. After the system has reached its asymptotic state under strain at $T = 35^{\circ}$C, we rapidly reheat it under the same strain up to $T = 55^\circ$C,  a value chosen so that stress relaxation vanish. We find that $\sigma$ recovers, within experimental error, the value $G_{T = 55^\circ}\times \epsilon$ corresponding to a purely elastic response at this temperature. 
Besides proving the absence of wall slip in our experiments at least up to the $50\%$ strain level used in this experiment, this result demonstrates the full reversibility of the physical process responsible for the observed stress evolution. 

\begin{figure}[htbp]
\begin{center}
\includegraphics[width = 8.5 cm]{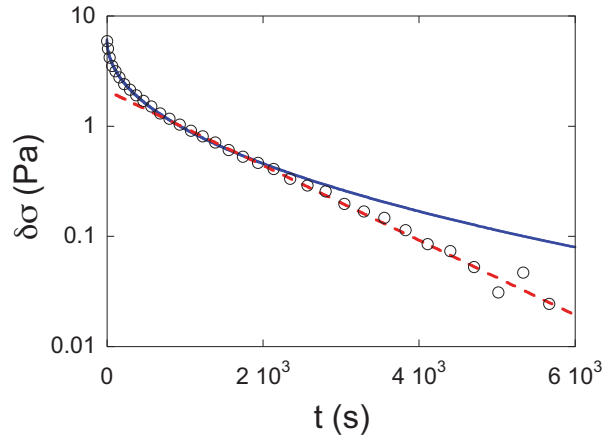}
\caption{Circles: Semi-logarithmic plot of $\delta\sigma(t) = \sigma(t) - \sigma_{\infty}$ for the data set of Fig.\ref{fig:relax}. Full line: Stretched exponential fit (see eq.\ref{eq:stretched}) of the initial decay, performed over $2\,10^3$  s and extrapolated to late times. Dashed line: asymptotic exponential decay extrapolated to early times. The fit parameters   are listed in the third line ($T = 35^\circ$C) of Table \ref{tab:fit}.  }
\label{fig:fits}
\end{center}
\end{figure}

We now turn to the analysis of the full time dependence of 
$$\delta\sigma(t) = \sigma(t) - \sigma_{\infty}$$
which characterizes relaxation towards the final, equilibrium state. As immediately appears from the semi-logarithmic plot of a typical set of data (see Fig.\ref{fig:fits}), the terminal decay of $\delta\sigma$ is a mere exponential, defining a final relaxation time $\tau_f$. However, this late stage fit does not account for the steeper  decrease observed at earlier times, which we find to be very well fitted by a stretched exponential of the form
\begin{equation}
\label{eq:stretched}
\delta\sigma(t) = \Delta\sigma \exp\left[-(t/\tau_{K})^{\beta}\right]
\end{equation}

These two regimes exhibit a rather narrow crossover about a time $\tau_{co}$ significantly larger than $\tau_{K}$.

The above described behavior is systematically observed, for $G = 1400$ Pa, for temperatures ranging from 45 to 25$^{\circ}$C. The corresponding values of the fit parameters are listed in Table \ref{tab:fit}, together with similar data obtained with a $G = 700$ Pa gel on a much narrower temperature range.

\begin{table}[htdp]
\caption{Fit parameters for the early and late relaxation regimes. }
\begin{center}
\begin{tabular}{|c|c|c|c|c|c|}
\hline
$G$&$T (^\circ$C)&$\beta$&$\tau_K$(s) &$\tau_{co}/\tau_K $&$\tau_f$(s) \\
\hline
1400&45&0.53&70&7.1&320\\
1400&40&0.50&170&3.5&680\\
1400&35&0.46&250&5.3&1850\\
1400&30&0.48&1850&3.7&7300\\
1400&25&0.41&3200&3.7&16500\\
\hline
700&40&0.34&920&3.1&5800\\
700&38&0.29&1600&2.5&10500\\
700&36&0.40&4100&2.3&16500\\
\hline
\end{tabular}
\end{center}
\label{tab:fit}
\end{table}

\section{Discussion}
\subsection{The relaxation process}
We first need to identify the physical process responsible for the observed stress relaxation. 

First of all, it is clear that it cannot be assigned to standard viscoelasticity. Indeed, from the values, on the order of 1 kPa, of their shear modulus, we can estimate the average mesh size of our gels $\xi \simeq (k_{B}T/G)^{1/3}$ to lie in the 10 nm range. The corresponding Rouse times $\tau_{R}$, on the order of $10^{-5}$s, are thus fully negligible on the relevant time scale, which ranges from hundreds to thousands of seconds. 

We must also exclude the relevance of the so-called ``slow mode'', observed in some scattering experiments and associated with frictional sliding of entanglements \cite{Shibayama}. Indeed, we have checked\cite{Supp} that the $5\%$ gelatin pre-gel solution which we use is in the semi-dilute, non entangled regime.
(Note, moreover, that the slow mode, when present, corresponds to a characteristic time of, typically, at most $10^{3} \tau_{R} \sim 10^{-2}$s). 
 
These remarks lead us to conclude that the stress decay triggered by a step strain in our system originates from a structural transformation which we identify as the strain-induced transition, briefly sketched out in Section II above, of at least part of the polymer strands from the coil to the helix configuration. In order to show that such a transition does indeed lead, following fast loading, to a stress {\it drop}, let us now very briefly summarize the simplified model formulated by Kutter and Terentjev \cite{Kutter}.

Consider a polymer strand, with fixed end-to-end distance $R$, containing N monomers of length $a$ arranged in two consecutive blocks, namely a helical segment, aligned with the end points, comprised of $n$ monomers, each of which occupies the effective length $\gamma a$ (with $\gamma < 1$) along the helix axis. The second block is a Gaussian coil of end-to-end distance $(R - \gamma na)$  formed by the remaining $(N-n)$ monomers. The strand free energy then reads:

\begin{equation}
\label {eq:energy}
F = n  \Delta f + \frac{3k_{B}T}{2(N-n)a^{2}} \left(R - \gamma na\right)^{2} + \Delta f_{int}
\end{equation}

The free energy gained per monomer in the helix (h) configuration $\Delta f = C(T - T_{0})$ vanishes linearly at the transition temperature $T_{0}$ of the free polymer chain. The second term in the r.h.s. is the elastic cost associated with imposing the end-to-end distance of the coiled block. The third one accounts for the presence one helix/coil and one helix/cross-link interfaces. For a given material (given value of $\gamma$), the number of monomers $n_{eq}$ engaged in the helix segment at equilibrium, obtained from $\partial F/\partial n = 0$, depends on the two dimensionless parameters: 

\begin{equation}
\label{eq:parameters}
x = \frac{R}{Na},\,\,\,\,\,\,\,\, \theta = \frac{2 \Delta f}{3k_{B}T} = \frac{2C(T - T_{0})}{3k_{B}T}
\end{equation} 

Since our experiments are performed at temperatures chosen to lie close above the transition temperature $T_{0} \simeq T_{gel}$ of the free chain, we will from now on specialize to the case where $\theta$ is a small positive number. 

\begin{figure}[htbp]
\begin{center}
\includegraphics[width = 7 cm]{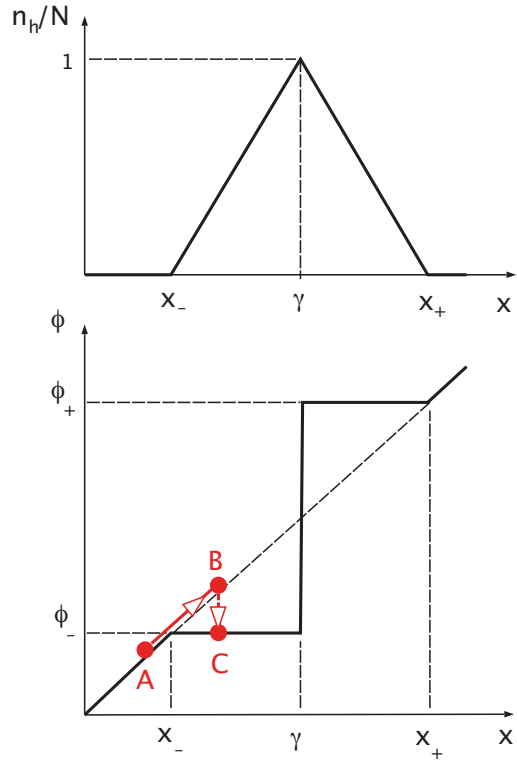}
\caption{Upper panel: Equilibrium helix fraction on a strand of $N$ monomers {\it vs.} reduced end-to-end distance $x=R/Na$. Lower panel:  Corresponding equilibrium strand tension $\phi$ (full black curve). The arrows correspond to the evolution of $\phi$ for the strain-induced path ABC (see text).      }
\label{fig:Kutter}
\end{center}
\end{figure}

The variation of $n_{eq}$ with $x$ at constant $T$ (fixed value of $\theta$) is shown on Fig.\ref{fig:Kutter}.a. Several regimes appear: at small extensions $x < x_{-} = \gamma - \sqrt{\gamma^{2} -\theta}$, no helix is present on the strand. Beyond this threshold, $n_{eq}$ grows linearly , until the full-helix state $n_{eq} = N$ is reached for $x = \gamma$. Further extension results in its symmetric linear decrease, up to the upper threshold $x_{+} =   \gamma + \sqrt{\gamma^{2} -\theta}$ above which the helix is completely unwound. As $T$ increases, the thresholds $x_{\pm}$ move toward $\gamma$, so that, for $\theta \geq  \gamma^{2}$, any strand is fully coiled whatever its extension. 

The tension $\phi_{eq} = \left(\partial F/\partial R\right)_{n_{eq}}$, shown on Fig.\ref{fig:Kutter}.b. For small $x < x_{-}$ it is given by the standard coil elastic form:  $\phi_{eq}^{c}(x) = 3k_{B}Tx/a$. In the (h/c) coexistence regime extending from $x_{-}$ to $\gamma$, it remains constant at the plateau value $\phi_{-} = \phi_{eq}^{c}(x_{-})$, then jumps to the upper plateau $\phi_{eq}^{c}(x_{+})$ in the unwinding coexistence regime, etc..    

Let us now sketch the evolution of a strand submitted to a step increase of its end-to-end extension, from the initial value $x_{A}$ to $x_{B} = x_{A}(1 + \epsilon)$. In order to fix ideas, we specialize to the case $x_{A} < x_{-}$, where the unstrained strand is a mere coil. Since the loading rate is very fast on the time scale $\tau_{c/h}$ of the isomerisation process needed for an elementary configurational change of order a few ten seconds), the evolution of the strand from points A to B (see Fig.\ref{fig:Kutter}.b) corresponding to the strain step occurs at the constant helix content $n_{eq}(x_{A})= 0$, where the tension increase is ruled by the instantaneous stiffness:

\begin{equation} 
\label{eq:stiffness}
\kappa = Na \left[\frac{\partial \phi}{\partial x}\right]_{n = n_{eq}(x_{A})} =  \frac{3k_{B}T}{Na^{2}}
\end{equation} 

Two cases are then possible.

$\bullet$ If $x_{B} > x_{-}$, $n_{eq}(x_{A}) \neq n_{eq}(x_{B})$, the strand configuration therefore relaxes until the helix content reaches its equilibrium value at the strained extension $x_{B}$, and the strand tension {\it decreases} from $\phi_{B}$ to $\phi_{C}$, the amplitude of this relaxation increasing with $(x_{B} - x_{-})$. 

$\bullet$ If, on the contrary, $x_{B} < x_{-}$, $n_{eq}(x_{B}) = n_{eq}(x_{A}) = 0$,   points B and C collapse and no retarded tension evolution occurs.

This sketch is easily extended to various other possible cases: for instance, if $n_{eq}(x_{B})$ and $x_{B}$ both lie in the $[x_{-}, \gamma]$ interval, after its step increase $\phi$ exhibits full relaxation to its unstrained value.

Let us now come back to our stress relaxation experiments. The gel random network contains gelatin strands with all possible orientations, and $R$ values distributed around the average mesh size $\bar{R}= \xi \sim 10$ nm. In order to evaluate its response, we have extended\cite{Supp} to the imposed simple shear deformation geometry the highly simplified network theory of ref.\cite{Kutter}, which assumes independent strands with a gaussian $x$-distribution, centered at $\bar{x} = \bar{R}/Na$, and of width $\Delta x$. 
Note that the measured fractional stress drop $\Delta\sigma/\sigma_{0}$ is small, typically of order a few percent. This indicates that, in the $T$-range of our interest, $\bar{x} + \Delta x \lesssim x_{-}$, i.e. that the helix content in the unstrained gel, carried by a tiny minority of strands in the tail of the $x$-distribution, is negligibly small. Strain drives beyond $x_{-}$ a larger fraction of the strands oriented close to the stretching principal axis, the evolution of which towards their equilibrium gives rise to the observed stress relaxation. As $T$ grows, so does the threshold $x_{-}$, and $\Delta\sigma/\sigma_{0}$ gradually vanishes, the blurring of the transition reflecting the distributed character of inter cross-link spacings.  This behavior (see Fig.4 of \cite{Supp}) is in qualitative agreement with the experimental stress drop results. 

This analysis, together with the full mechanical reversibility observed when reheating under constant strain a fully relaxed system, provides strong confirmation that the stress relaxation we observe does originate from the partial  transition to the helix configuration of stretched gelatin strands triggered by the applied strain.

\subsection{Relaxation dynamics}
Can we now understand the stress relaxation dynamics itself on the basis of the above framework - namely, as resulting from the evolution towards their equilibrium of a set of mechanically independent strained inter-crosslink strands? For this purpose, we need to build a dynamical, extended version of the static Kutter-Terentjev model \cite{Kutter}. On the basis of previous works on the helix-coil transition, we may reasonably assume \cite{Weaver} that the strand dynamics is a Fokker-Planck one, namely that the distribution $f(n,t \vert x)$ of the number of (h) monomers on a strand of dimensionless extension $x$ evolves according to:

\begin{equation}
\label{eq:FP}
\frac{\partial f}{\partial t} = D \frac{\partial }{\partial n}\left(\frac{\partial f}{\partial n} + \frac{1}{k_{B}T} \frac{\partial F(n,x)}{\partial n} f\right)
\end{equation}
where $F$ is defined by expression \ref{eq:energy} and the diffusion coefficient 

\begin{equation}
\label{eq:D}
D = \tau_{c/h}^{-1}
\end{equation}
is the inverse of the cis-trans isomerisation time.

One then immediately checks that the terminal relaxation of the average helix content $<n>$ from its initial value $n_{eq}(x_{i})$ to the strained equilibrium one $n_{eq}(x_{f} = x_{i}(1+\epsilon)$ is exponential with, in the case of interest here ($x{_i} < x_{-}$ and $x_{-} < x_{f} < \gamma>$), the characteristic time:

\begin{equation}
\label{eq:relaxstrand}
\tau_{strand} = \frac{k_{B}T}{D}\left[\frac{\partial^{2}F}{\partial n^{2}}\right]_{n_{eq}(x_{f})}^{-1} = \tau_{c/h} \frac{N(\gamma - x_{f})}{(\gamma^{2} - \theta)^{3/2}}
\end{equation} 
whose variation with temperature is primarily controlled by the Arrhenius dependence of $\tau_{c/h}$. As already mentioned, the associated activation energy  from existing biochemical results  $E_{act} \approx 0.75$ eV .

\begin{figure}[htbp]
\begin{center}
\includegraphics[width = 8.5 cm]{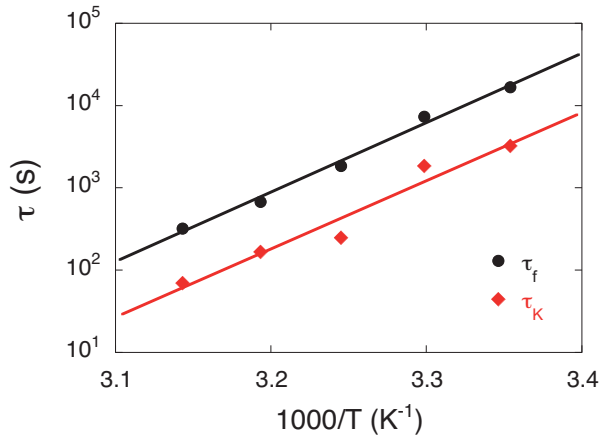}
\caption{Arrhenius plot of the Kolrausch time $\tau_K$  and of the terminal relaxation time $\tau_f$ for gels with $G = 1400$ Pa (data from Table \ref{tab:fit}). Lines: independent best exponential fits.   }
\label{fig:Arrhenius}
\end{center}
\end{figure}

This is to be compared with the measured stress relaxation time  $\tau_{f}$  of the gel. The Arrhenius plot of $\tau_{f}$  for the system with $ G = 1400$ Pa, shown on Fig.\ref{fig:Arrhenius},  is indeed linear over the (rather narrow) explored T-range. However, the associated energy 

\begin{equation}
\label{eq:activation}
{\cal E} = k_{B}\frac{d\left(\ln \tau_{f}\right)}{d\left(1/T\right)} = 1.7 \pm Ê 0.05\,\rm{ eV}
\end{equation}

turns out to be considerably larger than $E_{act}$. This discrepancy entails an important conclusion. Indeed, one could a priori be tempted to interpret the observed two-stage relaxation dynamics as resulting from independent activated jumps across energy barriers with a wide height distribution due to disorder. If such is the case, $\tau_f$ must be viewed as the activation time associated with the maximum barrier $\mathcal{E}$, of the form $\tau_f=\tau_0 \exp (\mathcal{E}/k_BT)$. Now, for $G = 1400$ Pa, we measure (see table~\ref{tab:fit}) $\tau_{f} = 7300$ s at $T = 30^{\circ}$C. At this temperature, $\exp \left( {\cal E}/k_{B}T\right) \sim 
10^{30}$, which would lead to an utterly unphysical value for the ``microscopic'' time prefactor, of order $10^{-26}$ s!

From this argument, we conclude that collective effects play a prominent part in the dynamics of our system.
    
Clearly, the dynamical version of the Kutter-Terentjev model misses an essential feature of the relaxation process. Indeed, the independent strand assumption overlooks an important physical point: consider an elementary (c) $\rightarrow$ (h) event occurring on the strand connecting nodes (i) and (j). Before the event, the tension forces on each node are equilibrated; the transition  induces a jump $\delta\phi_{ij}$ of the strand tension, i.e. a pair of extra forces $\pm\delta\phi_{ij}$ on (i) and (j). This force dipole results in an elastic deformation of the embedding network, and the corresponding long-ranged strain field, decaying as $r^{-3}$, in turn shifts  the end-to-end distance $R_{kl}$ of all other strands.  A given strand thus receives a set of  such signals, which constitute a self-generated "mechanical noise",  the effects of which combine with those of the thermal one to determine the relaxation dynamics.

This phenomenology is strongly reminiscent of the description of relaxation in deeply supercooled liquids in terms of local structural rearrangements and of the long range elastic Eshelby fields which they induce in the embedding medium \cite{PRLAnael}. Further support in favor of a close connection between deeply supercooled glass-formers and of our helix-forming gels is lent by the nature of the early stage stress relaxation, which we find to be unambiguously of the Kolrausch stretched exponential type ($\sim \exp[-(t/\tau_{K})^{\beta}]$) characteristic of glassy dynamics. 

Moreover, as shown on Fig.\ref{fig:Arrhenius}, $\tau_{K}$ shares with the terminal time $\tau_{f}$ an Arrhenius-like behavior, with the same anomalously large apparent activation energy ${\cal E}$. This we put in regard with the so-called fragility effect common to a majority of glass formers \cite{Dyre,Angell}  --- namely, upon approaching the glass transition from above, a faster than Arrhenius growth of the $\alpha$-relaxation time, which can be analyzed in terms of a growing apparent activation energy  (often fitted by the Vogel-Fulcher-Tamman expression ${\cal E} = AT/(T-T_0)$) which may become much larger than that for an elementary local event.

In our case, as discussed above, independent activated local events cannot account for the relaxation dynamics. This leads us to conclude that, in our system it is the combination of local relaxation events and of the elastic noise which they generate in the random network which is, most likely, responsible for the emergence of slow glassy dynamics.

Let us finally emphasize a conspicuous difference between our dilute random network and supercooled liquids, namely: in our gel, the Kolrausch regime unambiguously crosses over to a simple exponential decay with a relaxation time  $\tau_{f}$  proportional to $\tau_{K}$, while, in deeply supercooled glass-formers, no termination of the stretched exponential behavior has been observed  \cite {AngellMcK}.  

A hint about the origin of this difference could possibly be provided by the following remark. We have previously studied the aging behavior of gelatin networks \cite{Mixte} containing a hybrid population of thermoreversible and covalent crosslinks (CL). As is well known, the fully reversible gels exhibit a logarithmic growth of their shear modulus, which has not been observed to saturate \cite{Normand}, indicating that the upper limit of the relaxation spectrum, if any, lies far beyond times of order months. We have found that the presence of an increasing fraction of irreversible CL results in the exponential decrease of the "aging slope" $dG/d(\ln t)$. As compared with the triple helix physical CL, which are able to zip/unzip (hence to slide), covalent bonds prohibit the exchange of monomers between neighboring strands, thus reduce the possibility of monomer long range motion. We conjecture that this dynamical restriction constitutes the essential  difference between glass-formers and our covalent random networks. 

The tentative interpretation which we propose for our results clearly asks for quantitative tests, which  can only be provided by numerical studies. We note in this regard that, in contrast with reversible gels, thanks to their frozen CL topology, the fully covalent networks studied here appear amenable to realistic simulations.

\acknowledgments
We are grateful to M. Djabourov for an enlightening discussion. We thank Ana\"el Lema\^itre for a helpful critical reading of the manuscript.

\end{document}